\documentclass[aps,pra,groupedaddress,showpacs,twocolumn,superscriptaddress]{revtex4}
\usepackage{pstricks}
\usepackage{amsmath}
\usepackage{amsthm}
\usepackage{amsfonts}

\usepackage{graphics}

\def\bra#1{\mathinner{\langle{#1}|}}
\def\ket#1{\mathinner{|{#1}\rangle}}

\def\CP{C}
\def\dist{\delta}
\def\vect#1{\boldsymbol{#1}}
\def\vc{\vect{c}}
\def\olabel#1{}
\def\hsp{\hphantom{x}}

\def\ftna{\footnotemark[1]}
\def\ftnb{\footnotemark[2]}
\def\ftnc{\footnotemark[3]}
\def\ftnd{\footnotemark[4]}

\newcommand{\AC}{\mathcal{A}}
\newcommand{\CC}{\mathcal{C}}
\newcommand{\HC}{\mathcal{H}}
\newcommand{\PC}{\mathcal{P}}
\newcommand{\QC}{\mathcal{Q}}
\newcommand{\SC}{\mathcal{S}}
\newcommand{\UC}{\mathcal{U}}

\newcommand{\Tr}{{\rm Tr}}

\begin{document}

\title{Quantum Error Correcting Codes Using Qudit Graph States}
\author{Shiang Yong Looi}
\email[Electronic address:]{slooi@andrew.cmu.edu}
\affiliation{Department of Physics, Carnegie Mellon University, Pittsburgh,
Pennsylvania 15213, U.S.A.}

\author{Li Yu}
\affiliation{Department of Physics, Carnegie Mellon University, Pittsburgh,
Pennsylvania 15213, U.S.A.}

\author{Vlad Gheorghiu}
\affiliation{Department of Physics, Carnegie Mellon University, Pittsburgh,
Pennsylvania 15213, U.S.A.}

\author{Robert B. Griffiths}
\affiliation{Department of Physics, Carnegie Mellon University, Pittsburgh,
Pennsylvania 15213, U.S.A.}

\date{Version of 13 June 2008}

\begin{abstract} Graph states are generalized from qubits to collections of
  $n$ qudits of arbitrary dimension $D$, and simple graphical methods are used
  to construct both additive and nonadditive,  as well as degenerate and
  nondegenerate, quantum error correcting codes. Codes of distance 2 saturating
  the quantum Singleton bound for arbitrarily large $n$ and $D$ are constructed
  using simple graphs, except when $n$ is odd and $D$ is even.  Computer
  searches have produced a number of codes with distances 3 and 4, some
  previously known and some new.  The concept of a stabilizer is extended to
  general $D$, and shown to provide a dual representation of an additive graph
  code.
\end{abstract}

\maketitle

\section{Introduction\olabel{Introduction}}
\label{sct1}

Quantum error correction is an important part of various schemes for quantum
computation and quantum communication, and hence quantum error correcting
codes, first introduced about a decade ago \cite{shor,knill,steane} have
received a great deal of attention.  For a detailed discussion see Ch.~10 of
\cite{nielsen}. Most of the early work dealt with codes for qubits, with a
Hilbert space of dimension $D=2$, but qudit codes with $D>2$ have also been
studied \cite{rains3, ashikhmin, schlingemann, Schl02, Schl03, grassl2,
arvind}. They are of intrinsic interest and could turn out to be of some
practical value.

Cluster or graph states, which were initially introduced in connection with
measurement based or one-way quantum computing \cite{raussendorf}, are also
quite useful for constructing quantum codes, as shown in
\cite{schlingemann,Schl02,Schl03} in a context in which both the encoding
operation and the resulting encoded information are represented in terms of
graph states.  In the present paper we follow \cite{Schl03} in focusing on
qudits with general $D$, thought of as elements of the additive group
$\mathbb{Z}_D$ of integers mod $D$.  However, our strategy is somewhat
different, in that we use graph states and an associated basis (graph basis)
of the $n$-qudit Hilbert space in order to construct the coding subspace,
while \emph{not} concerning ourselves with the encoding process.  This leads
to a considerable simplification of the problem along with the possibility of
treating nonadditive graph codes on exactly the same basis as additive or
stabilizer codes.  It also clarifies the relationship (within the context of
graph codes as we define them) of degenerate and nondegenerate codes, though
in this paper we focus mainly on the latter.  The approach used here was
developed independently in \cite{cross} and \cite{yu2} for $D=2$, and in
\cite{hu} for $D>2$; thus several of our results are similar to those
reported in these references.

Following an introduction in Sec.~\ref{sct2} to Pauli operators, graph states,
and the graph basis, as used in this paper, the construction of graph codes is
the topic of Sec.~\ref{sct3}.  In Sec.~\ref{sct3a} we review the conditions
for an $(\!(n,K,\dist)\!)_D$ code, where $n$ is the number of carriers, $K$
the number of codewords or dimension of the coding space, $\dist$ the distance
of the code, and $D$ the dimension of the Hilbert space of one qudit.  We also
consider the distinction between degenerate and nondegenerate codes.  Our
definition of graph codes follows in Sec.~\ref{sct3b}, and the techniques we
use to find nondegenerate codes, which are the main focus of this paper, are
indicated in Sec.~\ref{sct3c}, while various results in terms of specific
codes are the subject of Sec.~\ref{sct4}.

In Sec.~\ref{sct4b} we show how to construct graph codes with $\dist=2$ that
saturate the quantum Singleton (QS) bound for arbitrarily large $n$ and $D$,
except when $n$ is odd and $D$ is even, and we derive a simple sufficient
condition for graphs to yield such codes.  For $n$ odd and $D=2$ we have an
alternative and somewhat simpler method of producing nonadditive codes of the
same size found in \cite{smolin}.  For both $D=2$ and $D=3$ we have studied
nondegenerate codes on sequences of cycle and wheel graphs, in
Secs.~\ref{sct4c} and \ref{sct4d}. These include a number of cases which
saturate the QS bound for $\dist=2$ and 3, and others with $\dist=3$ and 4
which are the largest possible additive codes for the given $n$, $D$, and
$\dist$.  Section~\ref{sct4d} contains results for a series of hypercube graphs
with $n=4$, 8, and 16, and in particular a $(\!(16,128,4)\!)_2$ additive code.

In Sec.~\ref{sct5} we show that what we call G-additive codes are stabilizer
codes (hence ``additive'' in the sense usually employed in the literature),
using a suitable generalization of the stabilizer formalism to general $D$.
In this perspective the stabilizer is a dual representation of a code which
is equally well represented by its codewords.
The final Sec.~\ref{sct6} has a summary of our results and indicates
directions in which they might be extended.

\section{Pauli operators and graph states\olabel{Graph states}}
\label{sct2}

\subsection{Pauli operators}
\label{sct2a}

Let $\{\ket{j}\}$, $j=0,1,\ldots D-1$ be an orthonormal basis for the
$D$-dimensional Hilbert space of a qudit, and define the unitary operators
\cite{ntk01}
\begin{equation}
\label{eqn1}
  Z = \sum_{j=0}^{D-1} \omega^j \ket{j} \! \bra{j},\quad
  X = \sum_{j=0}^{D-1} \ket{j} \! \bra{j \oplus 1},
\end{equation}
with $\oplus$ denoting addition mod $D$. They satisfy
\begin{equation}
\olabel{XZcommute}
\label{eqn2}
 Z^D=I=X^D, \quad XZ=\omega ZX,\quad
  \omega := \mathrm{e}^{2 \pi \mathrm{i} /D}.
\end{equation}
We shall refer to the collection of $D^2$ operators $\{X^\mu
Z^\nu\}$, $\mu,\nu=0,1,\ldots, D-1$, as (generalized) \emph{Pauli operators},
as they generalize the well known $I,X,Z,XZ\,(=-iY)$ for a qubit.  Together
they form the \emph{Pauli basis} of the space of operators on a qudit.

For a collection of $n$ qudits with a Hilbert space
$\HC=\HC_1\otimes\HC_2\otimes\cdots\HC_n$ we use subscripts to identify the
corresponding Pauli operators: thus $Z_l$ and $X_l$ operate on the space
$\HC_l$ of qudit $l$.  An operator of the form
\begin{equation}
\olabel{eqv1}
\label{eqn3}
 P = \omega^\lambda X^{\mu_1}_1 Z^{\nu_1}_1 X^{\mu_2}_2 Z^{\nu_2}_2
 \cdots X^{\mu_n}_n Z^{\nu_n}_n,
\end{equation}
where $\lambda$, and $\mu_l$ and $\nu_l$ for $1\leq l\leq n$, are integers in
the range $0$ to $D-1$, will be referred to as a \emph{Pauli product}.  If
$\mu_l$ and $\nu_l$ are both 0, the operator on qudit $l$ is the identity, and
can safely be omitted from the right side of \eqref{eqn3}.  The collection
$\QC$ of all operators $P$ of the form \eqref{eqn3} with $\lambda=0$, i.e., a
prefactor of 1, forms an orthonormal basis of the space of operators on $\HC$
with inner product $\langle A,\,B\rangle = D^{-n}\Tr(A^\dagger B)$; we call it
the (generalized) \emph{Pauli basis} $\QC$.

If $P$ and $Q$ are Pauli products, so is $PQ$, and hence the collection $\PC$
of all operators of the form \eqref{eqn3} for $n$ fixed form a multiplicative
group, the \emph{Pauli group}.  While $\PC$ is not Abelian, it has the
property that
\begin{equation}
\label{eqn4}
 PQ=\omega^\mu QP,
\end{equation}
where $\mu$ is an integer that depends on $P$ and $Q$.  (When $D=2$ and
$\omega=-1$ it is customary to also include in the Pauli group operators of
the form \eqref{eqn3} multiplied by $i$.  For our purposes this makes no
difference.)

The \emph{base} of an operator $P$ of the form \eqref{eqn3} is the collection
of qudits, i.e., the subset of $\{1,2,\ldots n\}$, on which the operator acts
in a nontrivial manner, so it is not just the identity, which is to say those
$j$ for which either $\mu_j$ or $\nu_j$ or both are greater than 0.  A general
operator $R$ can be expanded in the Pauli basis $\QC$, and its base is the
union of the bases of the operators which are present (with nonzero
coefficients) in the expansion.  The \emph{size} of an operator $R$ is defined
as the number of qudits in its base, i.e., the number on which it acts in a
nontrivial fashion.  For example, the base of $P=\omega^2 X^2_1 X^{}_4Z^{}_4$
(assuming $D\geq 3$) is $\{1,4\}$ and its size is 2; whereas the size of
$R=X^{}_1 + 0.5X^{}_2Z^2_2 Z^{}_3 +i X^{}_4$ is 4.

For two distinct qudits $l$ and $m$ the
\emph{controlled-phase} operation $\CP_{lm}$ on $\HC_l\otimes\HC_m$,
generalizing the usual controlled-phase for qubits, is
defined by
\begin{equation}
\olabel{CPdef}
\label{eqn5}
  \CP_{lm} = \sum_{j=0}^{D-1} \sum_{k=0}^{D-1}  \omega^{jk}
\ket{j} \! \bra{j} \otimes \ket{k} \! \bra{k}
= \sum_{j=0}^{D-1} \ket{j} \! \bra{j} \otimes Z^j_m.
\end{equation}
Of course, $C_{lm}=C_{ml}$, and it is easily checked that $(C_{lm})^D=I$.
It follows from its definition that $C_{lm}$ commutes with $Z_l$ and $Z_m$, and
thus with $Z_p$ for any qudit $p$.

\subsection{Graph states}
\label{sct2b}

\begin{figure}[ht]
\includegraphics{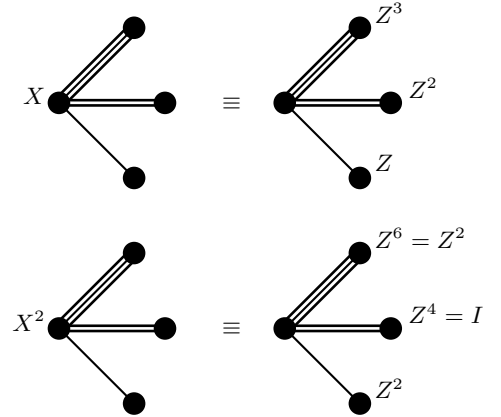}
\caption{Action of $X$ and $X^2$ on graph state ($D=4$).}
\olabel{fig:xonqudit}
\label{fgr1}
\end{figure}

Let $G=(V,E)$ be a graph with $n$ vertices $V$, each corresponding to a qudit,
and a collection $E$ of undirected edges connecting pairs of distinct vertices
(no self loops).  Multiple edges are allowed, as in Fig.\ref{fgr1} for
the case of $D=4$, as long as the multiplicity (weight) does not exceed $D-1$,
thus at most a single edge in the case of qubits.  The $lm$ element
$\Gamma_{lm}=\Gamma_{ml}$ of the \emph{adjacency matrix} $\Gamma$ is the
number of edges connecting vertex $l$ with vertex $m$.  The graph state
\begin{equation}
\label{eqn6}
 \ket{G} = \mathcal{U}\ket{G^0}= \mathcal{U}\left(\ket{+}^{\otimes n}\right),
\end{equation}
is obtained by applying the unitary operator
\begin{equation}
\olabel{bigU}
\label{eqn7}
 \mathcal{U} =
 \prod_{\{l,m\} \in E} \left(C_{lm}\right)^{\Gamma_{lm}}.
\end{equation}
to the product state
\begin{equation}
\label{eqn8}
 \ket{G^0} := \ket{+}\otimes\ket{+}\otimes\cdots\ket{+},
\end{equation}
where
\begin{equation}
  \ket{+} := D^{-1/2} \sum_{j=0}^{D-1} \ket{j}
\label{eqn9}
\end{equation}
is a normalized eigenstate of $X$, with eigenvalue 1.  In \eqref{eqn7} the
product is over all distinct pairs of qudits, with $(\CP_{lm})^0=I$ when $l$
and $m$ are not joined by an edge.  Since the $C_{lm}$ for different $l$ and
$m$ commute with each other, and also with $Z_p$ for any $p$, the order of the
operators on the right side of \eqref{eqn7} is unimportant.

Given the graph $G$ we define the \emph{graph basis} to be the set of
$D^n$ states
\begin{align}
\label{eqn10}
\ket{\vect{a}} &:= \ket{  a_1, a_2, \ldots ,a_n} = Z^{\vect{a}} \ket{G}
\notag\\
&= Z_1^{a_1}  Z_2^{a_2}  \cdots   Z_n^{a_n} \ket{G}
\end{align}
where $\vect{a}=(a_1,\ldots a_n)$ is an $n$-tuple of integers, each taking a
value between $0$ and $D-1$. The original graph state $\ket{G}$ is
$\ket{0,0,\ldots,0}$ in this notation.  That this collection forms an
orthonormal basis follows from the fact that the $Z_p$ operators commute with
the $\CP_{lm}$ operators, so can be moved through the unitary $\mathcal{U}$ on
the right side of \eqref{eqn6}.  As the states $Z^\nu\ket{+}$, $0\leq \nu \leq
D-1$, are an orthonormal basis for a single qudit, their products form an
orthonormal basis for $n$ qudits. Applying the unitary $\mathcal{U}$ to this
basis yields the orthonormal graph basis. The $n$-tuple representation in
\eqref{eqn10} is convenient in that one can define
\begin{align}
\label{eqn11}
  \ket{ \vect{a} \oplus \vect{b} } &:= \ket{  a_1 \oplus b_1, a_2 \oplus b_2,
  \ldots ,a_n \oplus b_n},
\notag\\
 \ket{ j  \vect{a}} &:= \ket{ j a_1, j a_2, \ldots ,j a_n  },
\end{align}
where $j$ is an integer between $0$ and $D-1$, and arithmetic operations are
mod $D$.

One advantage of using the graph basis is that its elements are mapped to each
other by a Pauli product (up to powers of $\omega$), as can be seen by
considering the action of $Z_l$ or $X_l$ on a single qudit.  The result for
$Z_l$ follows at once from \eqref{eqn10}.  And as shown in App.~\ref{sctpa} and
illustrated in Fig.~\ref{fgr1}, the effect of applying $X_l$ to $\ket{G}$ is
the same as applying $(Z_m)^{\Gamma_{lm}}$ to each of the qudits corresponding
to neighbors of $l$ in the graph.  Applying these two rules and keeping track
of powers of $\omega$ resulting from interchanging $X_l$ and $Z_l$,
see \eqref{eqn2}, allows one to easily evaluate the action of any Pauli
product on any $\ket{\vect{a}}$ in the graph basis.

\section{Code construction\olabel{Code}}
\label{sct3}

\subsection{Preliminaries}
\label{sct3a}

Consider a quantum code corresponding to a $K$-dimensional subspace,  with
orthonormal basis $\{ \ket{\vect{c}_q}\}$, of the Hilbert
space $\HC$ of $n$ qudits. When the Knill-Laflamme \cite{knill} condition
\begin{equation}
\olabel{eq:orthoOriginal}
\label{eqn12}
\bra{\vect{c}_q} Q \ket{\vect{c}_r} = f(Q) \delta_{qr}
\end{equation}
is satisfied for all $q$ and $r$ between $0$ and $K-1$, and every operator $Q$
on $\HC$ such that $1 \leq \mbox{size}(Q) < \dist$, but fails for some
operators of size $\dist$, the code is said to have \emph{distance} $\dist$,
and is an $(\!(n, K, \dist)\!)_D$ code; the subscript is often omitted when
$D=2$.  (See the definition of size in Sec.~\ref{sct2a}. The only operator of
size 0 is a multiple of the identity, so \eqref{eqn12} is trivially
satisfied.)  A code of distance $\dist$ allows the correction of any error
involving at most $\lfloor (\dist-1)/2\rfloor$ qudits, or an error on
$\dist-1$ (or fewer) qudits if the location of the corrupted qudits is already
known (e.g., they have been stolen).

It is helpful to regard \eqref{eqn12} as embodying two conditions: the obvious
off-diagonal condition saying that the matrix elements of $Q$ must vanish when
$r\neq q$; and the diagonal condition which, since $f(Q)$ is an arbitrary
complex-valued function of the operator $Q$, is nothing but the requirement
that all diagonal elements of $Q$ (inside the coding space) be identical.  The
off-diagonal condition has a clear analog in classical codes, whereas the
diagonal one does not. Both must hold for all operators of size up to and
including $\delta-1$, but need not be satisfied for larger operators.

In the coding literature it is customary to distinguish \emph{nondegenerate}
codes for which $f(Q)=0$ for all operators of size between 1 and $\delta-1$,
i.e., for \emph{all} $q$ and $r$
\begin{equation}
\olabel{eq:ortho}
\label{eqn13}
\bra{\vect{c}_q} Q \ket{\vect{c}_r} = 0 \;\; \text{for} \;\;
 1\leq\text{size}(Q)<\dist,
\end{equation}
and \emph{degenerate} codes for which $f(Q)\neq 0$ for at least one $Q$ in the
same range of sizes.  See p.~444 of \cite{nielsen} for the motivation behind
this somewhat peculiar terminology when $\delta$ is odd. In this paper our
focus is on nondegenerate codes. For the most part they seem to perform as well
as degenerate codes, though there are examples of degenerate codes that provide
a larger $K$ for given values of $n$, $\dist$, and $D$ than all known
nondegenerate codes. Examples are the $(\!(6, 2, 3)\!)_2$ \cite{ntk02} and
$(\!(25, 2, 9)\!)_2$ codes mentioned in \cite{calderbank}.

\subsection{Graph codes}
\label{sct3b}

When each basis vector $\ket{\vect{c}_q}$ is a member of the graph basis, of
the form \eqref{eqn10} for some graph $G$, we shall say that the corresponding
code is a \emph{graph code} associated with this graph.  As noted in
Sec.~\ref{sct1}, this differs from the definition employed in
\cite{schlingemann,Schl02,Schl03}, but agrees with that in more recent $D=2$
studies \cite{yu2, cross}, because we do not concern ourselves with the
processes of encoding and decoding.  In what follows we shall always assume
$\dist\geq 2$, since $\delta=1$ is trivial.  As the left side of \eqref{eqn12}
is linear in $Q$, it suffices to check it for appropriate operators drawn from
the Pauli basis $\QC$ as defined in Sec.~\ref{sct2a}.  It is helpful to note
that for any $Q\in\QC$, any pair $\ket{\vect{c}_q}$ and $\ket{\vect{c}_r}$ of
graph basis states and any $n$-tuple $\vect{a}$,
\begin{align}
\label{eqn14}
\bra{\vect{c}_q \oplus \vect{a} } Q \ket{\vect{c}_r \oplus \vect{a} } &=
\bra{\vect{c}_q} Z^{-\vect{a}} Q Z^{\vect{a}} \ket{\vect{c}_r}
\notag\\
 &=\omega^\mu\bra{\vect{c}_q} Q \ket{\vect{c}_r}
\end{align}
for some integer $\mu$ depending on $Q$ and $\vect{a}$; see \eqref{eqn10},
\eqref{eqn11} and \eqref{eqn4}.  Therefore, if \eqref{eqn12} is satisfied for
some $Q$ and a collection $\{\ket{\vect{c}_q}\}$ of codewords, the same will be
true for the same $Q$ and the collection $\{ \ket{\vect{c}_q \oplus \vect{a} }
\}$ (with an appropriate change in $f(Q)$).  Thus we can, and hereafter always
will, choose the first codeword to be
\begin{equation}
\label{eqn15}
 \ket{\vect{c}_0}=\ket{0,0,\ldots ,0}=\ket{G}.
\end{equation}

Analogous to Hamming distance in classical information theory we define
the \emph{Pauli distance} $\Delta$ between two graph basis states as
\begin{multline}
\label{eqn16}
\Delta(\vc_q,\vc_r) =\Delta \left( \ket{\vect{c}_q} ,\ket{\vect{c}_r} \right)
 :=\\ \min \{ \mbox{size($Q$)} : \bra{\vect{c}_q} Q \ket{\vect{c}_r} \neq 0 \},
\end{multline}
where it suffices to take the minimum for $Q\in\QC$, the Pauli basis.
(Ket symbols can be omitted from the arguments of $\Delta$ when the
meaning is clear.)
Also note the identities
\begin{align}
\olabel{eq:distIdentity}
\label{eqn17}
\Delta(\vc_q,\vc_r) &= \Delta(\vc_r,\vc_q) 
= \Delta(\vc_q \oplus \vect{a} ,\vc_r \oplus \vect{a}) \notag \\
&= \Delta(\vc_0,\vc_r\ominus\vc_q),
\end{align}
where $\vect{a}$ is any $n$-tuple, and $\ominus$ means difference mod $D$, see
\eqref{eqn11}.  The second equality is a consequence of \eqref{eqn14}.  Note
that if in \eqref{eqn16} we minimize only over $Q$ operations which are tensor
products of $Z$'s (no $X$'s), $\Delta$ is exactly the Hamming distance between
the $n$-tuples $\vect{c}_q$ and $\vect{c}_r$, see \eqref{eqn10}.

For the case $q=r$, where \eqref{eqn16} gives 0 (for $Q=I$), we introduce a
special \emph{diagonal distance} $\Delta'$ which is the minimum size of the
right side of \eqref{eqn16} when one restricts $Q$ to be an element of $\QC$
of size 1 or more.  The diagonal distance does not depend on the particular
value of $q=r$, but is determined solely by the graph state $\ket{G}$---see
\eqref{eqn14} with $r=q$---and thus by the graph $G$.  This has the important
consequence that if we consider a particular $G$ and want to find the optimum
codes for a given $\delta$ that is no larger than $\Delta'$, the collection of
operators $Q\in\QC$ for which \eqref{eqn12} needs to be checked will all have
zero diagonal elements, $f(Q)=0$, and we can use \eqref{eqn13} instead of
\eqref{eqn12}.  In other words, for the graph in question and for
$\delta\leq\Delta'$, all graph codes are nondegenerate, and in looking for an
optimal code one need not consider the degenerate case. Our computer results
in Sec.~\ref{sct4} are all limited to the range $\delta\leq\Delta'$ where no
degenerate codes exist for the graph in question. Any code with $\dist >
\Delta'$ will necessarily be degenerate, since there is at least
one nontrivial $Q$ for which \eqref{eqn12} must be checked for the diagonal
elements.

A code is \emph{G-additive (graph-additive)} if given any two codewords
$\ket{\vect{c}_q}$ and $\ket{\vect{c}_r}$ belonging to the code,
$\ket{\vect{c}_q \oplus \vect{c}_r}$ is also a codeword. As shown in
Sec.~\ref{sct5}, this notion of additivity implies the code is additive in the
sense of being a stabilizer code. For this reason, we shall omit the G in
G-additive except in cases where it is essential to make the distinction.
Codes that do not satisfy the additivity condition are called nonadditive.
The additive property allows one to express all codewords as ``linear
combinations'' of $k$ suitably chosen codeword generators. This implies an
additive code must have $K=D^r$, $r$ an integer, whenever $D$ is prime. We
will see an example of this in Sec.~\ref{sct4} for $D=2$.

The \emph{quantum Singleton} (QS) bound \cite{knill}
\begin{equation}
\olabel{eq:bound1}
\label{eqn18}
n \geq \log_D K + 2(\dist - 1) \;\;\; \text{or}
 \;\;\; K \leq D^{n - 2(\dist - 1)}
\end{equation}
is a simple but useful inequality.  We shall refer to codes which saturate
this bound (the inequality is an equality) as \emph{quantum Singleton} (QS)
codes. Some authors prefer the term MDS, but as it is not clear to us how
the concept of ``maximum distance separable,'' as explained in
\cite{macwilliams}, carries over to quantum codes, we prefer to use QS.

\subsection{Method}
\label{sct3c}

We are interested in finding ``good'' graph codes in the sense of a large $K$
for a given $n$, $\dist$, and $D$. The first task is to choose a graph $G$ on
$n$ vertices, not a trivial matter since the number of possibilities increases
rapidly with $n$.  We know of no general principles for making this choice,
though it is helpful to note, see App.~\ref{sctpa}, that the diagonal
distance $\Delta'$ cannot exceed 1 plus the minimum over all vertices of the
number of neighbors of a vertex.  Graphs with a high degree of symmetry are,
for obvious reasons, more amenable to analytic studies and computer searches
than those with lower symmetry.

Given a graph $G$ and a distance $\dist$, one can in principle search for the
best nondegenerate code by setting $\ket{\vc_0}=\ket{G}$, finding a
$\ket{\vc_1}$ with $\Delta(\vc_0,\vc_1)\geq \dist$, after that $\ket{\vc_2}$
with both $\Delta(\vc_0,\vc_2)\geq \dist$ and $\Delta(\vc_1,\vc_2)\geq \dist$,
and so forth, until the process stops.  However, this may happen before one
finds the largest $K$, because a better choice could have been made for
$\ket{\vc_q}$ at some point in the process. Exhaustively checking all
possibilities is rather time consuming, somewhat like solving an optimal
packing problem.

In practice what we do is to first construct a lookup table containing the
$D^n-1$ Pauli distances from $\ket{G}$ to all of the other graph basis states,
using an iterative process starting with all $Q\in\QC$ of size 1, then of size
2, etc. This process also yields the diagonal distance $\Delta'$.  As we are
only considering nondegenerate codes, we choose some $\dist\leq \Delta'$, so
that \eqref{eqn13} can be used in place of \eqref{eqn12}, and use the table to
identify the collection $S$ of all graph basis states with a distance greater
than or equal to $\dist$ from $\ket{\vc_0}=\ket{G}$.  If $S$ is empty there
are no other codewords, so $K=1$.  However, if $S$ is not empty then $K$ is at
least 2, and a search for the optimum code (largest $K$) is carried out as
follows.

We produce a graph $\SC$ (not to be confused with $G$) in which the nodes are
the elements of $S$, and an edge connects two nodes if the Pauli distance
separating them---easily computed from the lookup table with the help of
\eqref{eqn17}---is \emph{greater than or equal to} $\dist$. An edge in this
graph signifies that the nodes it joins are sufficiently (Pauli) separated to
be candidates for the code, and an optimal code corresponds to a largest
complete subgraph or \emph{maximum clique} of $\SC$.  Once a maximum clique has
been found, the corresponding graph basis states, including $\ket{\vc_0}$,
satisfy \eqref{eqn13} and span a coding space with the largest possible $K$ for
this graph $G$ and this $\dist$.

The maximum clique problem on a general graph is known to be NP-complete
\cite{garey} and hence computationally difficult, and we do not know if $\SC$
has special properties which can be exploited to speed things up.  We used the
relatively simple algorithm described in \cite{carraghan} for finding a
maximum clique, and this is the most time-consuming part of the search
procedure. 

The method just described finds additive as well as nonadditive codes. In fact
one does not know beforehand whether the resultant code will be additive or
not. If one is only interested in additive codes, certain steps can be modified
to produce a substantial increase in speed as one only has to find a set of
generators for the code.


\section{Results\olabel{Results}}
\label{sct4}

\subsection{Introduction}
\label{sct4a}

Results obtained using methods described above are reported here for various
sequences of graphs, each sequence containing graphs of increasing $n$ while
preserving certain basic properties.  We used a computer search to find the
maximum number $K$ of codewords for each graph in the sequence, for distances
$\dist\leq \Delta'$ and for $D=2$ or 3, qubits and qutrits, up to the largest
number $n$ of qudits allowed by our resources (running time). Sometimes this
revealed a pattern which could be further analyzed using analytic arguments or
known bounds on the number of codewords.

In the case of distance $\dist=2$ we can demonstrate the existence of QS codes
for arbitrarily large values of $n$ and $D$, except when $n$ is odd and $D$ is
even, see Part~A.  In the later subsections we report a significant collection
of $D=2$ and 3 codes for $\dist=2$, 3, and 4, including QS codes; codes which
are the largest possible additive codes for that set of $n$, $D$ and $\dist$;
and a new $(\!(16,128,4)\!)_2$ additive code.

Tables show the $K$ found as a function of other parameters. The meaning of
superscripts used in the tables is given below.
\begin{itemize}
\item $a$ -- Indicates the maximum clique search was terminated
before completion. This means the code we found might not be optimal, i.e.
there might be another code with larger $K$ for this graph. We can only say 
the code is \emph{maximal} in the sense that no codeword can be added without
violating \eqref{eqn13}. Absence of this superscript implies no code with a
larger $K$ exists for this $\dist$ and this graph, either because the program
did an exhaustive search, or because $K$ saturates a rigorous bound.
\item $b$ -- Indicates a nonadditive code. Codes without this superscript are
additive.
\item $c$ -- Indicates a QS code, one where $K$ saturates the Singleton bound
\eqref{eqn18}.
\item $d$ -- Indicates this is not a QS code, but the largest possible
  \emph{additive} (graph or other) code for the given $n$, $\dist$ and $D$,
  This follows from linear programming bounds in
  \cite{grassl1} for $D=2$ and \cite{ketkar} for $D=3$, along with the fact,
  Sec.~\ref{sct3b}, that for an additive code, $K$ must be an integer power
  of $D$ when $D$ is prime. A larger \emph{nonadditive} code for this graph
  might still be possible in cases flagged with $a$ as well as $d$. 

\end{itemize}

\subsection{Distance $\dist=2$; bar and star graphs}
\label{sct4b}

It was shown in \cite{calderbank} that for $D=2$ one can construct $\dist=2$
QS codes for any even $n$, and similar codes for larger $D$ are mentioned,
without giving details, in \cite{rains3}.  One way to construct graph codes
with $\dist=2$ is to use the method indicated in the proof, App.~\ref{sctpb},
of the following result.

\noindent
\textbf{Partition theorem}. \emph{Suppose that for a given $D$ the vertices of
  a graph $G$ on $n$ qudits can be partitioned into two nonempty sets $V_1$
  and $V_2$ with the property that for each vertex in $V_1$ the sum of the
  number of edges (the sum of the multiplicities if multiple edges are
  present) joining it to vertices in $V_2$ is nonzero and coprime to $D$, and
  the same for the number of edges joining a vertex in $V_2$ to vertices in
  $V_1$.  Then there is an additive QS code on $G$ with distance $\dist=2$.}

\begin{figure}
\includegraphics{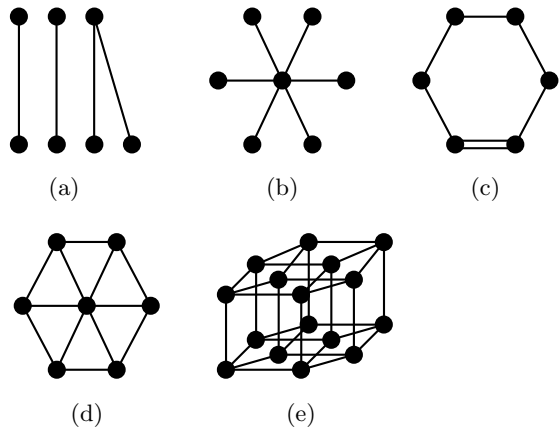}
\caption{Examples from different graph sequences: (a) bar (odd $n$), (b) star,
(c) cycle, (d) wheel, (e) $n=16$ hypercube.}
\label{fgr2}
\end{figure}

A \emph{bar} graph is constructed by taking $n$ vertices and dividing them
into two collections $V_1$ and $V_2$, of equal size when $n$ is even, and one
more vertex in $V_2$ when $n$ is odd, as in Fig.~\ref{fgr2}(a). Next pair the
vertices by connecting each vertex in $V_1$ by a single edge to a vertex in
$V_2$, with one additional edge when $n$ is odd, as shown in the figure.
(Multiple edges are possible for $D>2$, but provide no advantage in
constructing codes.)
When $n$ is even the conditions of the partition theorem are satisfied:
1 is always coprime to $D$.  For odd $n$, the last vertex in $V_1$ has
2 edges joining it to $V_2$, which is coprime to $D$ when $D$ is odd.  Hence
bar graphs yield $\dist=2$ QS codes for all $n$ when $D$ is odd, and for even
$n$ when $D$ is even.

A \emph{star} graph, Fig.~\ref{fgr2}(b), has a \emph{central} vertex joined by
single edges to every \emph{peripheral} vertex, and no edges connecting pairs
of peripheral vertices. Since the diagonal distance $\Delta'$ is 2,
nondegenerate star codes cannot have $\dist$ larger than 2.  As in the case of
bar codes, one can construct additive QS codes for any $n$ when $D$ is odd,
and for even $n$ when $D$ is even \cite{ntk04}. For odd $n$ and $D=2$ there
are nonadditive codes with
\begin{equation}
\label{eqn19}
K(n)=2^{n-2} - \frac{1}{2} \binom{n-1}{(n-1)/2};
\end{equation}
see App.~\ref{sctpc} for details.  Codes with these parameters were discovered
earlier by Smolin et al.\ \cite{smolin} using a different approach. Computer
searches show that for all odd $n\leq 7$ star graphs cannot yield a $K$ larger
than \eqref{eqn19}.

\subsection{Cycle graphs}
\label{sct4c}

We used computer searches to look for graph codes based on cycle (loop)
graphs, Fig.~\ref{fgr2}(c). Table~\ref{tbl1} shows the maximum number $K$ of
codewords for codes of distance $\dist=2$ and $\dist=3$ for both $D=2$ qubits
and $D=3$ qutrits. In the qutrit case the best codes were obtained by including
one double edge (weight 2), as in Fig.~\ref{fgr2}(c), though when $n$ is odd
equally good codes emerge with only single edges. In the qubit case all edges
have weight 1.

\begin{table}
\caption{Maximum $K$ for qubit and qutrit cycle graphs. See Sec.~\ref{sct4a}
for detailed meaning of superscripts.}
\olabel{tab:qubitCycle}
\label{tbl1}
\begin{tabular}{ c | c  c | c c }
\multicolumn{1}{c |}{ } &
\multicolumn{2}{c |}{$D=2$} &
\multicolumn{2}{c }{$D=3$}\\
$n$ &
\multicolumn{1}{c}{$\dist=2$} &
\multicolumn{1}{c|}{\hsp$\dist=3$\hsp} &
\multicolumn{1}{c}{$\dist=2$} &
\multicolumn{1}{c}{$\dist=3$} \\
\hline
4     & 4\ftnc         & 0             & 9\ftnc     & 1\ftnc   \\
5     & 6\ftnb         & 2\ftnc        & 27\ftnc    & 3\ftnc   \\
6     & 16\ftnc        & 1             & 81\ftnc    & 9\ftnc   \\
7     & 22\ftnb        & 2\ftnd        & 243\ftnc   & 27\ftnc  \\
8     & 64\ftnc        & 8\ftnd        & 729\ftnc   & 81\ftnc  \\
9     & 96\ftna\ftnb   & 12\ftnb       & 2187\ftnc  & 243\ftnc \\
10    & 256\ftnc       & 18\ftnb       & 6561\ftnc  & 729\ftnc \\
11    & 272\ftna\ftnb  & 32\ftna\ftnd  & 19683\ftnc & 729\ftna\ftnd \\
\hsp12\hsp    &\hsp1024\ftnc\hsp  &  \hsp64\ftna\ftnd\hsp
              &\hsp59049\ftnc\hsp &  \hsp2187\ftna\ftnd \hsp
\end{tabular}
\footnotetext[1]{Non-exhaustive search}
\footnotetext[2]{Nonadditive code}
\footnotetext[3]{Code saturating Singleton bound \eqref{eqn18}}
\footnotetext[4]{Largest possible additive code}
\end{table}

The $D=2$ entries in Table~\ref{tbl1} include for $n=5$ the well known $(\!(5,
2, 3)\!)_2$, the nonadditive $(\!(5, 6, 2)\!)_2$ presented in \cite{rains2},
and, for larger $n$, a $(\!(9, 12, 3)\!)_2$ code similar to that in \cite{yu}
and the $(\!(10, 18, 3)\!)_2$ of \cite{cross} based upon the same graph.

The $D=3$, $\dist=3$ entries are interesting because the QS bound is saturated
for $4 \leq n \leq 10$ but \emph{not} for $n = 11$. The $(\!(11, 3^6=729,
3)\!)_3$ code we found, the best possible \emph{additive} code according to
the linear programming bound in \cite{ketkar}, falls short by a factor of 3 of
saturating the $K=3^7=2187$ QS bound, and even a nonadditive code based on
this graph must have $K\leq 1990$ \cite{ntk03}.

One can ask to what extent the results for $\dist=2$ in Table~\ref{tbl1} could
have been obtained, or might be extended to larger $n$, by applying the
Partition theorem of Part A to a suitable partition of the cycle graph.  It
turns out---we omit the details---that when $D$ is odd one can use the
Partition theorem to produce codes that saturate the QS bound for any $n$, but
when $D$ is even the same approach only works when $n$ is a multiple of 4. In
particular, the $(\!(6, 16, 2)\!)_2$ additive QS code in Table~\ref{tbl1}
cannot be obtained in this fashion since the cycle graph cannot be partitioned
in the required way.

\subsection{Wheel graphs}
\label{sct4d}

If additional edges are added to a star graph so as to connect the peripheral
vertices in a cycle, as in Fig.~\ref{fgr2}(d), the result is what we call a
\emph{wheel} graph. Because each vertex has at least three neighbors, our
search procedure, limited to $\dist\leq \Delta'$, can yield $\dist=4$ codes on
wheel graphs, unlike cycle or star graphs.  The construction of $\dist=2$
codes for any $D$ is exactly the same as for star graphs, so in
Table~\ref{tbl2} we only show results for $\dist=3$ and 4, for both $D=2$ and
3. The $(\!(16, 128, 4)\!)_2$ additive code  appears to be new, and
its counterpart in the hypercube sequence is discussed below.

\begin{table}[ht]
\caption{Maximum $K$ for qubit and qutrit wheel graphs. See Sec.~\ref{sct4a}
for detailed meaning of superscripts.}
\olabel{tab:qubitCycle}
\label{tbl2}
\begin{tabular}{ c | c  c | c c }
\multicolumn{1}{c |}{ } &
\multicolumn{2}{c |}{$D=2$} &
\multicolumn{2}{c }{$D=3$}\\
$n$ &
\multicolumn{1}{c}{$\dist=3$} &
\multicolumn{1}{c|}{\hsp$\dist=4$\hsp} &
\multicolumn{1}{c}{$\dist=3$} &
\multicolumn{1}{c}{$\dist=4$} \\ \hline
6    & 1               & 1\ftnc        & 1                & 1\ftnc    \\
7    & 2\ftnd          & 0             & 27\ftnc          & 1         \\
8    & 8\ftnd          & 1\ftnd        & 27               & 9\ftnc    \\
9    & 8\ftnd          & 1\ftnd        & 243\ftnc         & 9         \\
10   & 20\ftnc         & 4\ftnd        & 243\ftna         & 27        \\
11   & 32\ftna\ftnd    & 4\ftnd        & 729\ftna\ftnd    & 81        \\
12   & 64\ftna\ftnd    & 8             & 2187\ftna\ftnd   & 81\ftna   \\
13   & 128\ftna\ftnd   & 16            & 6561\ftna\ftnd   & 243\ftna  \\
14   & 256\ftna\ftnd   & 32\ftna       & 19683\ftna\ftnd  & 729\ftna  \\
15   & 512\ftna\ftnd   & 64\ftna\ftnd  & \hsp59049\ftna\ftnd\hsp &
\hsp 2187\ftna\hsp \\
\hsp16\hsp  &\hsp1024\ftna\ftnd\hsp  & \hsp128\ftna\ftnd\hsp & &
\end{tabular}
\footnotetext[1]{Non-exhaustive search}
\footnotetext[2]{Nonadditive code}
\footnotetext[3]{Code saturating Singleton bound \eqref{eqn18}}
\footnotetext[4]{Largest possible additive code}
\end{table}

\subsection{Hypercube graphs}
\label{sct4e}

Hypercube graphs, Fig.~\ref{fgr2}(e), have a high symmetry, and as $n$
increases the coordination bound, App.~\ref{sctpa}, allows $\Delta'$ to
increase with $n$, unlike the other sequences of graphs discussed above.  We
have only studied the $D=2$ case, with the results shown in Table~\ref{tbl3}.
Those for $\dist=2$ are an immediate consequence of the Partition theorem:
each hypercube is obtained by adding edges between two hypercubes of the next
lower dimension, and these are the $V_1$ and $V_2$ of the theorem.  The
generators for the $(\!(16, 128, 4)\!)_2$ additive code are given in
Table~\ref{tbl4}. The $2^7=128$ codewords are of the form, see \eqref{eqn11},
$\ket{\alpha_1 \vect{g}_1\oplus \alpha_2 \vect{g}_2 \oplus \cdots \alpha_7
\vect{g}_7}$, where each $\alpha_j$ can be either 0 or 1.

\begin{table}[ht]
\caption{Maximum $K$ for qubit hypercube graphs. See Sec.~\ref{sct4a}
for detailed meaning of superscripts.}
\olabel{tab:qubitHypercube}
\label{tbl3}

\begin{tabular}{c|ccc}
\multicolumn{1}{c|}{} &
\multicolumn{3}{c}{$D=2$} \\
$n$ &
$\dist=2$ &  $\dist=3$ & \hsp$\dist=4$\hsp\\ \hline
4    & 4\ftnc     &  0        & 0 \\
8    & 64\ftnc    &  8\ftnd   & 1\ftnd \\
\hsp 16\hsp\hsp   & \hsp16384\ftnc\hsp &  512\ftna & 128\ftna\ftnd
\end{tabular}
\footnotetext[1]{Non-exhaustive search}
\footnotetext[3]{Code saturating Singleton bound \eqref{eqn18}}
\footnotetext[4]{Largest possible additive code}
\end{table}

\begin{table}
  \caption{Generators of  $(\!(16, 128, 4)\!)_2$ additive code for hypercube
graph}
\olabel{tab:hypercubeGen}
\label{tbl4}
\begin{tabular}{cc}
Generator & Bit notation \\
\hline
$\ket{\,\vect{g}_1}$ & \;\; $\ket{0,0,0,0, 0,0,0,0, 0,0,0,0, 1,1,1,1}$ \;\; \\
$\ket{\,\vect{g}_2}$ & $\ket{0,0,0,0, 0,0,0,0, 0,0,1,1, 0,0,1,1}$ \\
$\ket{\,\vect{g}_3}$ & $\ket{0,0,0,0, 0,0,0,0, 1,1,0,0, 0,0,1,1}$ \\
$\ket{\,\vect{g}_4}$ & $\ket{0,0,0,0, 0,0,1,1, 0,1,0,0, 0,1,0,0}$ \\
$\ket{\,\vect{g}_5}$ & $\ket{0,0,0,0, 1,1,0,0, 0,0,0,1, 0,0,0,1}$ \\
$\ket{\,\vect{g}_6}$ & $\ket{0,0,1,1, 0,0,0,0, 0,1,0,0, 0,1,0,0}$ \\
$\ket{\,\vect{g}_7}$ & $\ket{1,1,0,0, 0,0,0,0, 0,0,0,1, 0,0,0,1}$
\end{tabular}
\end{table}

\section{G-Additive Codes as Stabilizer Codes}
\olabel{stabilizer}
\label{sct5}

The stabilizer formalism introduced by Gottesman in \cite{gottesman} for $D=2$
(qubits) provides a compact and powerful way of generating quantum error
correcting codes. It has been extended to cases where $D$ is prime or a prime
power in \cite{ashikhmin, bahramgiri, ketkar}.  In \cite{Schl02} stabilizer
codes were extended in a very general fashion to arbitrary $D$ from a point of
view that includes encoding.  However, our approach to graph codes is somewhat
different, see Sec.~\ref{sct1}, and hence its connection with stabilizers
deserves a separate discussion. We will show that for any $D\geq 2$ a
G-additive (as defined near the end of Sec.~\ref{sct3b}) code is a stabilizer
code, and the stabilizer is effectively a dual representation of the code.

The Pauli group $\PC$ for general $n$ and $D$ was defined in Sec.~\ref{sct2a}.
Relative to this group we define a \emph{stabilizer} code (not necessarily a
graph code) $\CC$ to be a $K\geq 1$-dimensional subspace of the Hilbert space
satisfying three conditions:

\begin{description}

\item[C1.] There is a subgroup $\SC$ of $\PC$ such that
for \emph{every} $T$ in $\SC$ and \emph{every} $\ket{\psi}$ in $\CC$
\begin{equation}
\olabel{stabCondition1}
\label{eqn20}
 T \ket{\psi} = \ket{\psi}
\end{equation}

\item[C2.] The subgroup $\SC$ is maximal in the sense that every $T$ in $\PC$
  for which \eqref{eqn20} is satisfied for all $\ket{\psi}\in\CC$ belongs to
  $\SC$.

\item[C3.]  The coding space $\CC$ is maximal in the sense that any ket
  $\ket{\psi}$ that satisfies \eqref{eqn20} for every $T\in\SC$ lies in
  $\CC$.
\end{description}

If these conditions are fulfilled we call $\SC$ the \emph{stabilizer} of the
code $\CC$. That it is Abelian follows from \eqref{eqn4}, since for $K>0$
there is some nonzero $\ket{\psi}$ satisfying \eqref{eqn20}.  One can also
replace \eqref{eqn20} with
\begin{equation}
\label{eqn21}
 T \ket{c_q} = \ket{c_q}
\end{equation}
where the $\{\ket{c_q}\}$ form an orthonormal basis of $\CC$. Note that one
can always find a subgroup $\SC$ of $\PC$ satisfying C1 and C2 for any
subspace $\CC$ of the Hilbert space, but it might consist of nothing but the
identity. Thus it is condition C3 that distinguishes stabilizer codes from
nonadditive codes.  A stabilizer code is uniquely determined by $\SC$ as
well as by $\CC$, since $\SC$ determines $\CC$ through C3.

As we shall see, the stabilizers of G-additive graph codes can be described in
a fairly simple way.  Let us begin with one qudit, $n=1$, where the trivial
graph $G$ has no edges, and the graph basis states are of the form
$\{Z^c\ket{+}\}$ for $c$ in some collection $C$ of integers in the range
$0\leq c\leq D-1$. The subgroup $\SC$ of $\PC$ satisfying C1 and C2 must be of
the form $\{X^s\}$ for certain values of $s$, $0\leq s\leq D-1$, belonging to
a collection $S$.  This is because $Z$ and its powers map any state
$Z^c\ket{+}$ to an orthogonal state, and hence $T$ in \eqref{eqn21} cannot
possibly contain a (nontrivial) power of $Z$.  Furthermore, since
\begin{equation}
\label{eqn22}
  X^s Z^c \ket{+} = \omega^{cs}Z^c\ket{+},
\end{equation}
see \eqref{eqn2}, $X^s$ will leave $\{Z^c\ket{+}\}$ unchanged only if
$\omega^{cs}=1$, or
\begin{equation}
\label{eqn23}
cs \equiv 0 \pmod{D}.
\end{equation}
Thus for $\SC$ to satisfy C1, it is necessary and sufficient that
\eqref{eqn23} hold for every $c\in C$, as well as every $s\in S$. Further,
$\SC=\{X^s\}$ is maximal in the sense of C2 only if $S$ contains every $s$
satisfying \eqref{eqn23} for each $c\in C$. As shown in App.~\ref{sctpe}, such
a collection $S$ must either (depending on $C$) consist of $s=0$ alone, or
consist of the integer multiples $\nu s_1$, with $\nu=0,1,\ldots (D/s_1-1)$, of
some $s_1>0$ that divides $D$.  In either case, $S$ is a subgroup of the group
$\mathbb{Z}_D$ of integers under addition mod $D$, and indeed any such subgroup
must have the form just described.

We now take up C3. Given the maximal collection $S$ of solutions to
\eqref{eqn23}, we can in turn ask for the collection of $C'$ of integers $c$
in the range $0$ to $D-1$ that satisfy \eqref{eqn23} for every $s$ in $S$.
Obviously, $C'$ contains $C$, but as shown in App.~\ref{sctpe}, $C'=C$ if and
only if $C$ is a subgroup of $\mathbb{Z}_D$, i.e., $\CC$ is G-additive. Next
note that every $T$ in $\SC$, as it is a power of $X$ and because of
\eqref{eqn22}, maps every graph basis state to itself, up to a phase. Thus when
(and only when) $\CC$ is G-additive, the codewords are just those graph basis
states for which this phase is 1 for every $T\in\SC$.  To check C3, expand an
arbitrary $\ket{\psi}$ in the graph basis. Then $T\ket{\psi}=\ket{\psi}$ for
all $T\in\SC$ means that all coefficients must vanish for graph basis states
that do not belong to $\CC$. Hence C3 is satisfied if and only if $\CC$ is
G-additive.

The preceding analysis generalizes immediately to $n>1$ in the case of the
trivial graph $G^0$ with no edges.  A graph code $\CC$ has a basis of the form
$\{Z^{\vect{c}}\ket{G^0}\}$ for a collection $C$ of integer $n$-tuples
$\vect{c} \in \mathbb{Z}_D^n$, and is G-additive when the collection
$C=\{\vect{c}\}$ is closed under component-wise addition mod $D$, i.e., is a
subgroup of $\mathbb{Z}_D^n$. Whether or not $\CC$ is G-additive, the subgroup
$\SC$ of $\PC$ satisfying C1 and C2 consists of all operators of the form
$X^{\vect{s}}=X_1^{s_1}X_2^{s_2}\cdots$ with the $n$-tuple $\vect{s}$
satisfying
\begin{equation}
\label{eqn24}
\vect{c}\cdot\vect{s} := \sum_{l=1}^n c_l s_l \equiv \vect{0} \pmod{D}
\end{equation}
for every $\vect{c}\in C$.  Just as for $n=1$, $\SC$ cannot contain Pauli
products with (nontrivial) powers of $Z$ operators. Let $S$ denote the
collection of all such $\vect{s}$. The linearity of \eqref{eqn24} means $S$ is
an additive subgroup of $\mathbb{Z}_D^n$.

One can also regard \eqref{eqn24} as a set of conditions, one for every
$\vect{s}\in S$, that are satisfied by certain $\vect{c}\in\mathbb{Z}_D^n$.
The set $C'$ of all these solutions is itself an additive subgroup of
$\mathbb{Z}_D^n$, and contains $C$.  In App.~\ref{sctpe} we show that $C'=C$
if and only if $C$ (the collection we began with) is an additive subgroup
of $\mathbb{Z}_D^n$, and when this is the case the sizes of $C$ and $S$ are
related by
\begin{equation}
\label{eqn25}
 |C|\cdot|S| = D^n.
\end{equation}
Just as for $n=1$, any $X^{\vect{s}}$ maps a graph basis state for the trivial
graph $G^0$---they are all product states---onto itself up to a multiplicative
phase, and the same argument used above for $n=1$ shows that C3 is satisfied
for all $T\in\SC$ if and only if $\CC$ is G-additive.

To apply these results to a general graph $G$ on $n$ qubits, note that the
unitary $\UC$ defined in \eqref{eqn7} provides, through \eqref{eqn6} and
\eqref{eqn10}, a one-to-one map of the graph basis states of the trivial $G^0$
onto the graph basis states of $G$. At the same time the one-to-one map $\UC P
\UC^\dagger$ carries the $\SC$ satisfying C1 and C2 (and possibly C3) for the
$G^0$ code to the corresponding $\SC$, satisfying the same conditions for the
$G$ code. (The reverse maps are obtained by interchanging $\UC^\dagger$ and
$\UC$.)  Consequently, the results obtained for $G^0$ apply at once to $G$,
and the transformation allows the elements of the stabilizer for the $G$ graph
code to be characterized by integer $n$-tuples $\vect{s}$ satisfying
\eqref{eqn24}. Thus we have shown that G-additive codes are stabilizer codes,
and for these the coding space and stabilizer group descriptions are dual,
related by \eqref{eqn24}: each can be derived from the other.

\section{Conclusion and Discussion}
\olabel{conclusion}
\label{sct6}

In this paper we have developed an approach to graph codes which works for
qudits with general dimension $D$, and employs graphical methods to search for
specific examples of such codes.  It is similar to the approaches developed
independently in \cite{cross,yu2,hu}.  We have used it for computer searches
on graphs with a relatively small number $n$ of qudits, and also to construct
certain families of graphs yielding optimum distance $\dist=2$ codes for
various values of $D$ and $n$ which can be arbitrarily large.  It remains a
challenging problem to do the same for codes with distance $\dist>2$.

In a number of cases we have been able to construct what we call quantum
Singleton (QS) codes that saturate the quantum Singleton bound \cite{knill}:
these include the $\dist=2$ codes for arbitrarily large $n$ and $D$ mentioned
above, and also a number of $\dist=3$ codes in the case of $D=3$ (qutrits),
see Tables~\ref{tbl1} and \ref{tbl2}.  The results for cycle graphs for $D=3$
and $\dist=3$ in Table~\ref{tbl1} are interesting in that the QS bound is
saturated for $n\leq 10$, but fails for $n=11$, as it must for nondegenerate
codes; see the discussion in Sec.~\ref{sct4c}.  Our results are consistent
with the difficulty of finding QS codes for larger $\dist$ \cite{grassl1}, but
suggest that increasing $D$ may help, as observed in \cite{schlingemann}.  It
is worth noting that we have managed to construct many of the previously known
nonadditive codes, or at least codes with the same $(\!(n,K,\dist)\!)_D$,
using simple graphs. Some other nonadditive codes not discussed here, such as
the $(\!(10, 24, 3)\!)_2$ code in \cite{yu2}, can also be obtained from
suitably chosen graphs.  While all these results are encouraging, they
represent only a beginning in terms of understanding what properties of graphs
lead to good graph codes, and how one might efficiently construct such codes
with arbitrarily large $n$ and $\dist$, for various $D$.

As noted in Sec.~\ref{sct3b}, all graph codes with distance $\dist\leq
\Delta'$, where $\Delta'$ is the diagonal distance of the graph, are
necessarily nondegenerate, and our methods developed for such codes will (in
principle) find them all.  All codes with $\dist> \Delta'$ are necessarily
degenerate codes, and their systematic study awaits further work.
It should be noted that our extension of graph codes to $D>2$ is based on
extending Pauli operators in the manner indicated in \cite{hostens}.  Though
the extension seems fairly natural, and it is hard to think of alternatives
when $D$ is prime, there are other ways to approach the matter when $D$ is
composite (including prime powers), which could yield larger or at least
different codes, so this is a matter worth exploring.

The relationship between stabilizer (or additive) codes and G-additive
(as defined in Sec.~\ref{sct3b}) graph codes has been clarified by showing
that they are dual representations, connected through a simple equation,
\eqref{eqn24}, of the same thing.  One might suspect that such duality extends
to nongraphical stabilizer codes, but we have not studied the problem outside
the context of graph codes. Nonadditive codes, which---if one uses our
definition, Sec.~\ref{sct5}---do not have stabilizers, are sometimes of larger
size than additive codes, so they certainly need to be taken into account in
the search for optimal codes. The graph formalism employed here works in
either case, but computer searches are much faster for additive codes.

\begin{acknowledgments}
The research described here received support from the National Science
Foundation through Grant No. PHY-0456951. The authors would like to thank
Markus Grassl and Bei Zeng for very helpful comments and discussions.
\end{acknowledgments}

\appendix
\section{The X-Z rule and related}
\olabel{app:xzRule}
\label{sctpa}

\noindent
\textbf{X-Z Rule}.  \emph{Acting with an $X$ operator on the $i'th$ qudit of a
  graph state $\ket{G}$ produces the same graph basis state as the action of
  $Z$ operators on the neighbors of qudit $i$, raised to the power given by
  the edge multiplicities $\Gamma_{im}$.}

The operator $X_i$ commutes with $\CP_{lm}$ when $i \neq l$ and $i \neq m$, but
if $i=l$ (or similarly $i=m$) one can show using \eqref{eqn5} and \eqref{eqn1}
that
\begin{equation}
\olabel{apxCeqn2}
\label{Aeqn1}
X_{l} \CP_{lm}  =  \CP_{lm} Z_m X_l = Z_m \CP_{lm} X_{l}.
\end{equation}
That is, an $X_i$ operator can be pushed from left to right through a
$\CP_{lm}$ with at most the cost of producing a $Z$ operator associated with
the \emph{other} qudit: if $i=l$ one gets $Z_m$, if $i=m$ one gets $Z_l$.
Since all $Z$ commute with all $\CP$, one can place the resulting $Z_m$ either
to the left or to the right of $\CP_{lm}$.

Now consider pushing $X_i$ from the left to the right through $\mathcal{U}$,
the product of $\CP_{lm}$ operators defined in \eqref{eqn7}. Using
\eqref{Aeqn1} successively for those $\CP_{lm}$ that do not commute with
$X_i$, one sees that this can be done at the cost of generating a $Z_m$ for
every edge of the graph connecting $i$ to another vertex $m$. Let the product
of these be denoted as $\hat Z:= \prod_{(l=i,m) \in E}Z_{m}^{\Gamma_{lm}}$.
Then, with definition \eqref{eqn6}, we can show
\begin{align}
\label{Aeqn2}
X_i \ket{G} &= X_i \mathcal{U} \ket{G^0} = \hat Z \mathcal{U} X_i \ket{G^0}
\notag\\
 &= \hat Z\mathcal{U}\ket{G^0} =\hat Z \ket{G},
\end{align}
which completes the proof of the X-Z Rule.

For graph codes satisfying \eqref{eqn13}, the X-Z Rule leads to the:

\noindent
\textbf{Coordination bound}. \emph{The diagonal distance $\Delta'$ for a graph
  $G$ cannot exceed $\nu+1$, where $\nu$ is the minimum over all vertices of
  the number of neighbors of a vertex, this being the number of vertices
  joined to the one in question by edges, possibly of multiplicity greater
  than 1.}

To make the counting absolutely clear consider Fig.~\ref{fgr1}, where the
vertex on the left has 3 neighbors, and each of the others has 1 neighbor, so
that in this case $\nu=1$.  To derive the bound, apply $X$ to a vertex which
has $\nu$ neighbors.  By the X-Z rule the result is the same as applying
appropriate powers of $Z$ to each neighbor.  Let $P$ be this $X$ tensored with
appropriate compensating powers of $Z$ at the neighboring vertices in such a
way that $P\ket{G}=\ket{G}$. The size of $P$ is $\nu+1$, and $\Delta'$ can be
no larger.  \smallskip

Another useful result follows from the method of proof of the X-Z Rule:

\noindent
\textbf{Paulis to Paulis}. \emph{Let $P$ be a Pauli product \eqref{eqn3},
  and for $\UC$ defined in \eqref{eqn7} let
\begin{equation}
\label{Aeqn3}
P' = \UC^\dagger P \UC,\quad P'' = \UC P \UC^\dagger.
\end{equation}
Then both $P'$ and $P''$ are Pauli products.}

To see why this works, rewrite the first equality as $\UC P' = P \UC$, and
imagine pushing each of the single qudit operators, of the form
$X_j^{\mu_j}Z_j^{\nu_j}$, making up the product $P$ through $\UC$ from left to
right. This can always be done, see the discussion following \eqref{Aeqn1}, at
the cost of producing some additional $Z$ operators, which can be placed on
the right side of $\UC$, to make a contribution to $P'$.  At the end of the
pushing the final result can be rearranged in the order specified in
\eqref{eqn3} at the cost of some powers of $\omega$, see \eqref{eqn2}.  The
argument for $P''$ uses pushing in the opposite direction.

\section{Partition theorem proof}
\olabel{app:bipartite}
\label{sctpb}

Given the partition of the $n$ qudits into sets $V_1$ and $V_2$ containing
$n_1$ and $n_2$ elements, the code of interest consists of the graph basis
states $\ket{ \vect{c}} = \ket{c_1, c_2, \ldots,c_n}$ satisfying the two
conditions
\begin{eqnarray}
\sum_{i \in V_1} c_i & \equiv & 0 \pmod{D}
\olabel{eq:parity1}
\label{Beqn1}
\\
\sum_{j \in V_2} c_j & \equiv & 0 \pmod{D}
\olabel{eq:parity2}
\label{Beqn2}
\end{eqnarray}
This code is additive and contains $K=D^{n_1-1} \times D^{n_2-1} = D^{n-2}$
codewords. (The counting can be done by noting that \eqref{Beqn1} defines a
subgroup of the additive group $\mathbb{Z}_D^{n_1}$, and its cosets are
obtained by replacing 0 with some other integer on the right side of
\eqref{Beqn1}.)

We first demonstrate that this code has $\dist\geq 2$ by showing that any
Pauli operator, except the identity, applied to a single qudit maps a codeword
into a graph basis state not in the code. If $Z^\nu$ for $0 < \nu < D$ is
applied to a qudit in $V_1$, the effect will be to replace 0 on the right side
of \eqref{Beqn1} with $\nu$, so this graph state is not in the code. If
$X^\mu$, $0 < \mu < D$ is applied to a qudit in $V_1$ the result according to
the X-Z Rule, App.~\ref{sctpa}, will be the same as placing $Z$ operators on
neighboring qudits in $V_2$ (as well as $V_1$) in such a way that 0 on the
right side of \eqref{Beqn2} is replaced by $g\mu$, where $g$ is the total
number of edges (including multiplicities) joining the $V_1$ qudit with qudits
in $V_2$.  But as long as $g$ is coprime to $D$, as specified in the condition
for the theorem, $g\mu$ cannot be a multiple of $D$, and \eqref{Beqn2} will no
longer be satisfied.  The same is true if $Z^\nu X^\mu$ is a applied to a
qudit in $V_1$.  Obviously the same arguments work for Pauli operators applied
to a single qudit in $V_2$. Thus we have shown that $\dist\geq 2$.

But $\dist>2$ is excluded by the QS bound, so we conclude that we have an
additive code of $K=D^{n-2}$ elements and distance $\dist=2$ that saturates
the QS bound.

\section{Construction of qubit star graph codes\olabel{app:qubitStarGraph}}
\label{sctpc}

As noted in Sec.~\ref{sct4b} a star graph for $n$-qubits consists of a central
vertex joined by edges to $n-1$ peripheral vertices.  Let $V_1$ be the
central vertex and $V_2$ the set of peripheral vertices.  When $n$ is even and
$D=2$ the conditions of the Partition theorem, Sec.~\ref{sct4b}, are
satisfied, and the $\dist=2$ code constructed in App.~\ref{sctpb} consists of
the $2^{n-2}$ graph basis states with no $Z$ on the central qubit and an even
number $r$ of $Z$'s on the peripheral qubits, thus satisfying \eqref{Beqn1}
and \eqref{Beqn2}, and yielding an additive QS code.

When $n$ is odd the central vertex is connected to an even number $n-1$ of
vertices in $V_2$, so the conditions of the Partition theorem no longer hold.
A reasonably large $\dist=2$ nonadditive code can, however, be constructed by
again assuming no codeword has $Z$ on the central qubit, and that the code
contains all graph basis states with $r$ $Z$'s on the peripheral qubits
\emph{for a certain selected set $R$ of $r$ values}.

The set $R$ must satisfy two conditions. First, it cannot contain both $r$ and
$r+1$, because applying an additional $Z$ to a codeword with $r$ $Z$'s yields
one with $r+1$, and one cannot have both of them in a code of distance
$\dist=2$.  Second, applying $X$ to the central vertex and using the
X-Z rule, App.~\ref{sctpa}, maps a codeword with $r$ $Z$'s to one with
$r'=n-1-r$; hence $R$ cannot contain both $r$ and $n-1-r$.  For example, when
$n=7$ ($n-1=6$ peripheral qubits) the set $R=\{0,2,5\}$ satisfies both
conditions, as does $R=\{1,4,6\}$, whereas $R=\{1,2,6\}$ violates the first
condition and $R=\{1,3,5\}$ the second.

By considering examples of this sort, and noting that the number of such graph
basis states with $r$ $Z$'s is $\binom{n-1}{r}$ which is equal to
$\binom{n-1}{n-1-r}$, one sees that for $n$ odd one can construct in this way a
nonadditive code with
\begin{equation}
\label{Ceqn1}
\sum_{i=0}^{(n-3)/2} \binom{n-1}{i} =
 2^{n-2}  - \frac{1}{2} \binom{n-1}{(n-1)/2}
\end{equation}
codewords.

\section{Solutions to $\vect{c}\cdot\vect{s} \equiv 0 \pmod{D}$
\olabel{app:Gadditive}}
\label{sctpe}

Let $\AC$ be the collection of all $n$-component integer vectors (i.e.,
$n$-tuples) of the form $\vect{a} = (a_1,a_2,\ldots a_n)$,
$0\leq a_j \leq D-1$, with component-wise sums and scalar multiplication
defined using arithmetic operations mod $D$. In particular, $\AC$ is a group
of order $D^n$ under component-wise addition mod $D$.  We shall be
interested in subsets $C$ and $\SC$ of $\AC$ that satisfy
\begin{equation}
  \vect{c}\cdot\vect{s} := \sum_{l=1}^n c_l s_l \equiv 0 \pmod{D}
\label{Eeqn1}
\end{equation}
for all $\vect{c}\in C$ and $\vect{s}\in S$.  Given some collection $C$,
we shall say that $S$ is \emph{maximal} relative to $C$ if it includes
\emph{all} solutions $\vect{s}$ that satisfy \eqref{Eeqn1} for every
$\vect{c}\in C$. It is easily checked that a maximal $S$ is an additive
subgroup of $\AC$: it includes the zero vector and $-\vect{s}$ mod $D$ whenever
$\vect{s}\in S$.  A similar definition holds for $C$ being maximal relative to
a given $S$. We use $|C|$ to denote the number of elements in a set or
collection $C$.

\textbf{Theorem}.  \emph{Let $C$ be an additive subgroup of $\AC$, and let
  $S$ be maximal relative to $C$, i.e., the set of all $\vect{s}$ that
  satisfy \eqref{Eeqn1} for every $\vect{c}\in C$.  Then $C$ is also
  maximal relative to $S$, and
\begin{equation}
  |C|\cdot |S| = D^n.
\label{Eeqn2}
\end{equation}}

The proof is straightforward when $D$ is a prime, since $\mathbb{Z}_D$ is a
field, and one has the usual rules for a linear space. The composite case is
more difficult, and it is useful to start with $n=1$:

\textbf{Lemma}.
  Let $C$ be a subgroup under addition mod $D$ of the integers lying between
  $0$ and $D-1$, and $S$ all integers in the same range satisfying
\begin{equation}
  cs \equiv 0 \pmod{D}
\label{Eeqn3}
\end{equation}
for every $c\in C$. Then $C$ consists of \emph{all} integers $c$ in the
range of interest which satisfy \eqref{Eeqn3}, and $|C|\cdot |S|= D$.

When $C=\{0\}$ the proof is obvious, since $|C|=1$ and $|S|=D$.  Otherwise,
because it is an additive subgroup of $\mathbb{Z}_D$, $C$ consists of the
multiples $\{\mu c_1\}$ of the smallest positive integer $c_1$ in $C$,
necessarily a divisor of $D$, when $\mu$ takes the values $0, 1, \ldots
s_1-1$, where $s_1=D/c_1$.  One quickly checks that all integer multiples
$s=\nu s_1$ of this $s_1$ satisfy \eqref{Eeqn3} and are thus contained in $S$.
But $S$ is also an additive subgroup, and $s_1$ is its minimal positive
element (except in the trivial case $c_1=1$), for were there some smaller
positive integer $s'$ in $S$ we would have $0<c_1s' <D$, contradicting
\eqref{Eeqn3}.  Similarly there is no way to add any additional integers to
$C$ while preserving the subgroup structure under addition mod $D$ without
including a positive $c$ less than $c_1$, which will not satisfy \eqref{Eeqn3}
for $s=s_1$.

For $n>1$ it is helpful to use a \emph{generator matrix} $F$, with components
$F_{rl}$, each between $0$ and $D-1$, with the property that $\vect{c}\in C$
if and only if it can be expressed as linear combinations of rows of $F$,
i.e.,
\begin{equation}
  c_l \equiv \sum_r b_r F_{rl} \pmod{D}
\label{Eeqn4}
\end{equation}
for a suitable collection of integers $\{b_r\}$. This collection will of course
depend on the $\vect{c}$ in question, and for a given $\vect{c}$ need not be
unique, even assuming (as we shall) that $0\leq b_r\leq D-1$.  In particular
the matrix $F$ for which each row is a distinct $\vect{c}$ in $C$, with $r$
running from $1$ to $|C|$, is a generator matrix.  It is straightforward to
show that if $F$ is any generator matrix for $C$, $S$ consists of all solutions
$\vect{s}$ to the equations
\begin{equation}
  \sum_{l=1}^n F_{rl} s_l \equiv 0 \pmod{d} \text{ for } r=1, 2, \ldots.
\label{Eeqn5}
\end{equation}

The collections $C$ and $S$, vectors of the form \eqref{Eeqn4} and those
satisfying \eqref{Eeqn5}, remain the same if $F$ is replaced by another
generator matrix $F'$ obtained by one of the following \emph{row operations}:
(i) permuting two rows; (ii) multiplying (mod $D$) any row by an
\emph{invertible} integer, i.e., an integer which has a multiplicative inverse
mod $D$; (iii) adding (mod $D$) to one row an \emph{arbitrary} multiple (mod
$D$) of a different row; (iv) discarding (or adding) any row that is all
zeros, to get a matrix of a different size. Of these, (i) and (iv) are
obvious, and (ii) is straightforward. For (iii), consider what happens if the
second row of $F$ is added to the first, so that $F'_{rl}=F^{}_{rl}$ except
for
\begin{equation}
  F'_{1l} \equiv F^{}_{1l}+ F^{}_{2l} \pmod{D}.
\label{Eeqn6}
\end{equation}
Then setting
\begin{equation}
  b'_1=b_1,\; b'_2\equiv b_2-b_1\pmod{d},\; b'_l=b_l \text{ for } l\geq 3
\label{Eeqn7}
\end{equation}
leads to the same $\vect{c}$ in \eqref{Eeqn4} if $b$ and $F$ are replaced by
$b'$ and $F'$ on the right side. Likewise, any $\vect{c}$ that can be written
as a linear combination of $F'$ rows can be written as a combination of those
of $F$, so the two matrices generate the same collection $C$, and hence have
the same solution set $S$ to \eqref{Eeqn5}.  Since adding to one row a
different row can be repeated an arbitrary number of times, (iii) holds for
an arbitrary (not simply an invertible) multiple of a row.

The corresponding column operations on a generator matrix are (i) permuting
two columns; (ii) multiplying a column by an invertible integer; (iii) adding
(mod $D$) to one column an arbitrary multiple (mod $D$) of a different column.
Throwing away (or adding) columns of zeros is \emph{not} an allowed operation.
When column operations are carried out to produce a new $F'$ from $F$, the new
collections $C'$ and $S'$ obtained using \eqref{Eeqn4} and \eqref{Eeqn5}
will in general be different, but $C'$ is an additive subgroup of the same
size (order), $|C'|=|C|$, and likewise $|S'|=|S|$. The argument is
straightforward for (i) and (ii), and for (iii) it is an easy exercise to show
that if the second column of $F$ is added to the first to produce $F'$, the
collection $C$ is mapped into $C'$ by the map
\begin{equation}
 c_1' \equiv c_1+c_2 \pmod{D}\: ;\quad c'_l = c_l \text{ for } l\geq 2
\label{Eeqn8}
\end{equation}
whose inverse will map $C'$ into $C$ when one generates $F$ from $F'$ by
subtracting the second column from the first. Thus $|C|= |C'|$.
The same strategy shows that $|S'|=|S|$; instead of \eqref{Eeqn8}
use $s'_2\equiv s_2-s_1\pmod{D}$, and $s'_l=s_l$ for $l\neq 2$.

The row and column operations can be used to transform the generator matrix to
a (non unique) diagonal form, in the following fashion.  If each $F_{rl}$ is
zero the problem is trivial.  Otherwise use row and column permutations so
that the smallest positive integer $f$ in the matrix is in the upper left
corner $r=1=l$. Suppose $f$ does not divide some element, say $F_{13}$, in the
first row.  Then by subtracting a suitable multiple of the first column from
the third column we obtain a new generator $F'$ with $0<F'_{13}<f$, and
interchanging the first and third columns we have a generator with a smaller,
but still positive, element in the upper left corner.  Continue in this
fashion, considering both the first row and the first column, until the upper
left element of the transformed generator divides \emph{every} element in
both. When this is the case, subtracting multiples of the first column from
the other columns, and multiples of the first row from the other rows, will
yield a matrix with all zeros in the first row and first column, apart from
the nonzero upper left element at $r=1=l$, completing the first step of
diagonalization.

Next apply the same overall strategy to the sub matrix obtained by ignoring the
first row and column.  Continuing the process of diagonalization and
discarding rows that are all zero (or perhaps adding them back in again), one
arrives at a diagonal $n\times n$ generator matrix
\begin{equation}
  \hat F_{rl} = f_l\delta_{rl},
\label{Eeqn9}
\end{equation}
where some of the $f_l$ may be zero.  The counting problem is now much
simplified, because for each $l$ $c_l$ can be any multiple mod $D$ of $f_l$,
and $s_l$ any solution to $f_l s_l\equiv 0 \pmod{D}$, independent of what
happens for a different $l$.  Denoting these two collections by $C_l$ and
$S_l$, the lemma implies that $|C_l|\cdot|S_l|=D$ for every $l$, and taking
the product over $l$ from $1$ to $n$ yields \eqref{Eeqn2}.  This in turn
implies that $C$ consists of \emph{all possible} $\vect{c}$ that satisfy
\eqref{Eeqn1} for all the $\vect{s}\in S$. To see this, note that the size
$|C|$ of $C$ is $D^n/|S|$.  If we interchange the roles of $C$ and $S$ in the
above argument (using a generator matrix for $S$, etc.), we again come to the
result \eqref{Eeqn2}, this time interpreting $|C|$ as the number of solutions
to \eqref{Eeqn1} with $S$ given.  Thus since it cannot be made any larger, the
original additive subgroup $C$ we started with is maximal relative to $S$.
This completes the proof.


\begin{thebibliography}{31}
\bibitem[{ntk({\natexlab{a}})}]{ntk01}
\bibinfo{note}{See \cite{hostens} for a list of references to work that employs
  operators of this type.}
\bibitem[{ntk({\natexlab{b}})}]{ntk02}
\bibinfo{note}{While there seems to be no proof that the $(\!(6, 2, 3)\!)_2$
  degenerate code has a larger $K$ than any nondegenerate code with $n=6$ and
  $\dist=3$, some support comes from the fact that we performed an exhaustive
  search of all graphs with 6 vertices and did not find a nondegenerate graph
  code with $\dist=3$ and $K>1$. But the notion that this degenerate code is
  superior to nondegenerate codes is undercut by the observation that the well
  known nondegenerate $(\!(5, 2, 3)\!)_2$ code uses only 5 instead of 6 qubits
  to achieve equal values of $K$ and $\dist$.}
\bibitem[{ntk({\natexlab{c}})}]{ntk04}
\bibinfo{note}{We omit the details. In some but not all cases one can use the
  Partition theorem with $V_1$ and $V_2$ the center and the peripheral
  vertices. Allowing some double edges when $D>2$ extends the range of $n$
  values where the Partition theorem can be employed.}
\end{thebibliography}

\end{document}